# A study of the magnetotransport properties of the graphene (II. Fractional Quantum Hall Effect)

M. A. Hidalgo

Departamento de Física y Matemáticas
Universidad de Alcalá
Alcalá de Henares, Madrid, Spain
Correspondence and request for materials should be addressed to miguel.hidalgo@uah.es

**Abstract**

We present an approach to the fractional quantum Hall effect observed in grapheme (GFQHE), basing us on the model developed previously for the fractional quantum Hall effect in a two-dimensional electron system embedded in a quantum well (FQHE) [Hidalgo, 2013(*)]. The main idea in the view proposed f0r the FQHE is the breaking of the symmetry of the Hamiltonian of every electron in a two-dimensional electron gas (2DEG) under the application of a magnetic field and in the presence of an electrostatic potential due to the ionized impurities. As the magnetic field increases the effect of that electrostatic potential evolves; changing in turn the spatial symmetry of the Hamiltonian: from continuous to discrete one. The model provides the odd fractional states, and corresponding plateaux, $p/3$, $p$ being any integer, observed in graphene.

(*) *All the references from the author can be downloaded from his "Researchgate" website*

## Introduction

The FQHE is known since more than thirty decades ago. After the pristine measurements by Tsui et al., [1982], who observed the appearance of the 1/3 plateau, the importance of the phenomenon were confirmed with the observation in any two dimensional electron system

(2DES) embedded in quantum well semiconductors (QW), of large series of other fractional plateaux, most of them odd: {4/5, 2/3, 3/5, 4/7, 5/9, 5/3, 8/5, 11/7, 10/7, 7/5, 4/3, 9/7} [Clark et al., 1986]; {5/3, 8/5, 10/7, 7/5, 4/3, 9/7, 4/5, 3/4, 5/7, 2/3, 3/5, 4/7, 5/9, 6/11, 7/13, 6/13, 5/11, 4/9, 3/7, 2/5, and also 8/3, 19/7, 33/13, 32/13, 7/3, 16/7}[Willet et al., 1987, 1988]; {2/3, 7/11, 3/5, 4/7, 5/9, 6/11, 5/11, 4/9, 3/7, 2/5, 1/3, 2/7, 3/11, 4/15, 3/13, 2/9} [Du et al., 1988]; {14/5, 19/7, 8/3, 13/5, 23/9, 22/9, 17/7, 12/5, 7/3, 16/7, 11/5} [Choi et al., 2007]; {19/5, 16/5, 14/5, 8/5, 7/3, 11/5, 11/3, 18/5, 17/5, 10/3, 13/5, 12/5} [Shabani et al., 2010]. Although sometimes, but with much less frequency, even planteaux: {15/4, 7/2, 13/4, 11/4, 5/2, 9/4} [Clark et al., 1986]; {11/4, 5/2, 9/4} [Willet et al., 1987]; {1/4} [Willet et al., 1988]; {11/4, 21/8, 5/2, 19/8, 9/4} [Du et al., 1988]; {7/2, 5/2} [Choi et al., 2007].

The phenomenon shows their main features in the data related to magnetotransport in a two-dimensional electron system (2DES): minima or zeroes in the longitudinal resistance, i.e. Shubnikov-de Haas effect (SdH), and well-defined plateaux in the Hall resistance (at the fractional values of the fundamental Hall resistance $R_H = h/e^2$ ).

Concerning on its physical explanation, the FQHE was interpreted in the beginning as a consequence of the formation of correlated states in the electron liquid behavior of the 2DES, [Laughlin, 1983]. In fact, Laughlin proposed that the origin of the observed FQHE 1/3, -as well any $1/q$, with $q$ being an odd integer-, is due to the formation of a correlated incompressible electron liquid; and from that view the electron-electron interaction is analyzed constructing several electron wave-functions *ad hoc*, i.e. therefore, theoretically all the objective in analyzing the phenomenon is searching for explicit trial wave functions corresponding to the states of the 2DES that does not break any continuous spatial symmetry and show energy gaps. The nature of these states is related to uniform density condensates. Because after the discovery of the 1/3 plateau, many other fractional plateaux have been observed, following the general expression $p/(2sp\pm1)$, with $s$ and $p$ being integers, the initial idea by Laughlin has more recently been elaborated, assuming the appearance of new composite quasiparticles of fractional electronic charges, coming from the combination of electrons and flux quanta, the so-called composite fermions, according to which the FQHE may be viewed as an integer quantum Hall effect (IQHE) of quasiparticles consisting of an electron capturing an even number of flux quanta [Jain, 1992]. These composite entities become elementary particles of the system.

Analyzing all the experiments related to FQHE on different 2DESs, the fundamental character of the FQHE phenomenon is obvious, seeming clear the common origin of the physics underneath. (In this context, it is important to highlight the fact that in all FQHE measurements the integer plateaux are also observed.)

In fact, a completely different approach to the FQHE has recently been proposed, [Hidalgo, 2013], in which the phenomenon appears to be a physical consequence of the effect of the interaction of electrons with impurities in the quantum limit condition.

On the other hand, in a previous work, [Hidalgo, 2014], the authors present an alternative single electron approach to the integer quantum Hall effect in monolayer graphene (GIQHE), natural extension of the model developed for the integer quatum Hall effect (IQHE) in QW semiconductor, [Hidalgo, 1995, 1998, 2007]. It reproduces all the observed characteristics of the magnetotransport properties in the monolayer graphene (MLG), and integrating it in the same framework as the IQHE and the FQHE: the same first physical principles, hypothesis and assumptions. The only different aspect between both two dimensional electron systems, QW and graphene, would be the quantization of two bands, the VB and CB, in that last case.

The fractional quantum Hall effect in graphene (GFQHE), also appear to be so fundamental as the FQHE in QW. However, the theoretical explanation is nowadays an open question.

The present work wants to be a simple theoretical approach to the problem, extension of our previous model for the FQHE in QW, [Hidalgo, 2013], to the GFQHE.

**Origin of fractional odd denominator quantum levels in graphene**

MLG is a flat layer of carbon atoms arranged in a hexagonal lattice with two carbon atoms per unit cell. Of the four valence states, three $sp^2$ orbitals form a $\sigma$ state with three neighbouring carbon atoms, and one $p$ orbital develops into delocalized $\pi$ and $\pi^*$ states that provide the occupied valence band (VB) and the lowest unoccupied conduction band (CB). We now consider $\varepsilon_{VB}$ and $\varepsilon_{CB}$ as the corresponding Hilbert spaces associated with the dynamic states of each band, valence and conduction. We assume the same effective masses, i.e., $m^*_{VB} = m^*_{CB} = m^*$, where $m^*_{VB}$ and $m^*_{CB}$ would correspond to the effective masses of the valence and conduction bands, respectively.

As we have already mentioned above, our approach to the GFQHE will be based on the same assumptions and single electron approach as the IQHE and FQHE in QW semiconductor, [Hidalgo 1995, 2013]. Thus, our first task will be to determine the quantized electron states, taking into account the symmetric gauge $\vec{A} = B(-y,x)/2$, being $\vec{B} = (0,0,B)$ the magnetic field, assumed to be perpendicular to the plane defined by the graphene. In the present case, the resultant Hilbert space for the dynamical states of any electron under the application of the magnetic field will be the tensor product of the dynamical spaces of each band, i.e., $\varepsilon^B = \varepsilon^B_{VB} \otimes \varepsilon^B_{CB}$, where $\varepsilon^B_{VB}$ and $\varepsilon^B_{CB}$ are the corresponding dynamical spaces for the valence and conduction bands under the presence of $B$.

Therefore, due to the contribution of the two bands, the electrons will have two degrees of freedom and, then, the quantized energy spectrum under the application of $B$ has to be described by that corresponding to an isotropic bidimensional harmonic oscillator. Hence, the quantized levels will be given by the expression

$$E_n = (i+j)\hbar\omega_0 = n\hbar\omega_0 = nE_0 \tag{1}$$

with $i, j$, natural numbers related to each dynamical space, $n=0, 1, 2...$, $\omega_0 = eB/m^*$ the fundamental angular frequency, and $E_0 = \hbar\omega_0$. These levels are degenerate in all possible angular momentum states, determined by $m$. Another consequence of $\varepsilon^B = \varepsilon^B_{VB} \otimes \varepsilon^B_{CB}$, is that the module of the wave number has now to be

$$k = \sqrt{k^{CB} + k^{VB}}$$

On the other hand, we know that the expected value of the square of the distance from the center of the trajectory of the electron to the origin -in our case each ionized impurities, Figure 1-, for each band, referenced in all below as Larmor radius, $R_O$, is given by

$$<R_O^2> = qR^2 \tag{2}$$

being $q$ an odd number, and $R = \sqrt{\hbar/eB}$, the magnetic wavelength. Hence, we can define the wavelength associated with this Larmor radius as

$$\lambda_q = \lambda_q^{CB} = \lambda_q^{VB} = 2\pi\sqrt{<R_O^2>} \tag{3}$$

(that is expected to be the same for both bands). Thus, we have now for the case of graphene

$$k_q = \sqrt{k_q^{CB} + k_q^{VB}} = \sqrt{2}k_q^{VB} \tag{4}$$

that can be written as

$$k_q = \sqrt{2}\frac{2\pi}{\lambda_q} = \sqrt{\frac{2eB}{q\hbar}} \tag{5}$$

It is well-known that the general Hamiltonian of every electron in the 2DEG is

$$H = \frac{1}{2m^*}(\vec{p}+e\vec{A})^2 + U(\vec{r}) = \frac{1}{2m^*}(\vec{p}+e\vec{A})^2 + U^i(\vec{r}) + U^e(\vec{r}) \tag{6}$$

The last two terms correspond to the energy contribution of the electrostatic potentials due to the ionized impurities, $U^i(\vec{r})$, and the electron-electron interaction, $U^e(\vec{r})$. (Later we will discuss about the Zeeman term.) Of course, we consider that the electrostatic potentials are perturbations respect to the effect of the magnetic field over the electrons.

Now, below, we summarize the model, although a detail description of it can be found in Hidalgo [2013].

At low magnetic field the Larmor radius of every electron is large; then their interactions $U^i(\vec{r})$ with the ionized impurities, because in that case we have $d_i \ll 2R_O$, $R_O$ given by Equation (2) and $d_i$ being the distance among ionized impurities, can be assumed to be a uniform term, and thus, the energy states for every electron are $E_n = E_0 + U_0$. (The electrostatic potential associated with the electron-electron interaction can be neglected in average due to the symmetry related to the electron distribution in the 2DES.)

However, at high magnetic fields the Larmor radius will be of the order of $d_i$, and, then, the effect of the ionized impurities over every electron will now contribute to the Hamiltonian with a non-uniform term $U^i(\vec{r})$, what involves a change in the symmetry of the Hamiltonian of the 2DEG, leaded from the continuous spatial symmetry at low magnetic field to a discrete one at high magnetic fields. Therefore, under this last condition, we can view the new states as an arrangement of cyclotron orbits according to the expected symmetry of the distribution of the ionized impurities, Figure 1. To characterize this new symmetry we introduce a correlation length related to that short-range order, which we will express as $\eta d_i$. Then, this new symmetry implies that $H(r) = H(r + \alpha \eta d_i)$, where $\alpha$ is a natural number.

Taking into account the Landau functions, $\{\psi_n^m\}$, associated with the cyclotron orbits, we can express their spatial correlation through the relation, [Hidalgo, 2013],

$$<\psi_n^m(r+\alpha\eta d_i)/H_2/\psi_n^m(r+[\alpha\pm 1]\eta d_i)>=\pm\gamma/2 \tag{7}$$

where we suppose that $\gamma$ is the same for all cyclotron orbits. (We also assume the higher correlations terms negligible.) The Landau functions does not verify the Bloch theorem, but we can construct a new base reflecting that new short range order taking the linear combination of the cyclotron orbits functions, [Hidalgo, 2013].

Under all these conditions, the new energy states can be expressed by the equation

$$E = E_n \mp \gamma \pm \frac{\gamma}{2}(k_q\eta d_i)^2 \tag{8}$$

from where it is easy to deduce that $\gamma = U_0$. Then, if we write $\hat{E}_n = E_n \mp U_0$, we arrive to the relation

$$E = \hat{E}_n \pm \frac{\gamma}{2}(k_q\eta d_i)^2 = \hat{E}_n + \hat{E}_q^\eta \tag{9}$$

with $\hat{E}_q^\eta = \pm\eta^2\frac{U_0}{2}(k_q d_i)^2$. Considering the effective mass $m^* = \frac{\hbar^2}{|U_0|d_i^2}$, and using (5) we can write

$$\hat{E}_q^\eta = \pm\frac{\eta^2}{q}E_0 \tag{10}$$

Hence, the energy states are

$$E = \hat{E}_n \pm \frac{\eta^2}{q}E_0 = nE_0 \pm \frac{\eta^2}{q}E_0 + U_0 = \left(n\pm\frac{\eta^2}{q}\right)E_0 + U_0 \tag{11}$$

As we have described above, the expected correlations in the arrangement of the electron cyclotron orbits in the 2DES in graphene is like that shown in Figure 1, short range order to the first neighbors, which correspond to values of the correlation length, $d_i$, i.e., $\eta=1$. Therefore, we have the energy states for every electron

$$\frac{E}{\hbar\omega_0} = \left(n\pm\frac{1}{q}\right) \tag{12}$$

In Table I we detail the values of Equation (12) for the Landau levels *n*=0, 1 and 2, and *q*=3. As it is seen, the values obtained coincide with the family of odd plateaux observed in the experiments of the GFQHE, [Bolotin et al., 2009], [Du et al., 2009], [Gharari et al., 2011], [Dean et al., 2011].

Finally, the contribution of the Zeeman and the spin-orbit coupling terms is summarized in the expression $E_{spin} = g^* m^* \hbar \omega_0 / 4m$, being *g*\* the generalized gyromagnetic factor. In the FQHE (and also GFQHE) conditions we can consider that all the electrons are uniformly polarized.

To illustrate the results of the model we will show the simulation of both magnetoconductivities (SdH and Hall effects), for any 2DEG in graphene under fractional quantum Hall conditions. Again, the detail description of the general procedure can be found in Hidalgo, [2013]. Here we only summarize it.

The density of states for *q* is then given by

$$g_q(E) = g_0 \left\{ 1 + 2 \sum_{p=1}^{\infty} A_{\Gamma,p} cos\left[ \left( \frac{2\pi q p E}{\hbar \omega_0} - \frac{g^*}{4} \frac{m^*}{m} \right) \right] \right\} \tag{13}$$

where $A_{\Gamma,p} = exp\left\{ -\frac{2\pi^2 p^2 \Gamma^2}{\hbar^2 \omega_c^2} \right\}$, is the gaussian term related to the width of the energy levels, and due to the interaction of electrons with defects and impurities, [Ando et al., 1982]. (For sake of simplicity we have assumed gaussian width for each energy level, independent of the magnetic field.)

From the density of sates the electron density is easily obtained, [Hidalgo, 1995, 1998],

$$n = n_0 + \delta n = n_0 + \frac{2eB}{hq} \sum_{p=1}^{\infty} \frac{1}{\pi p} A_{\Gamma,p} sen\left[ X_{q,F} - \frac{g^*}{4} \frac{m^*}{m} \right] \tag{14}$$

with $X_{q,F} = 2\pi q p\, E_F/\hbar\omega_0$. (In this expression we have supposed very low temperature, the ideal experimental condition to observe the FQHE.) $E_F$ is the Fermi level of the 2DEG, given by

$$E_F = \frac{2\pi\hbar^2 n_e}{m^*} \qquad (15)$$

where $n_e$ is the electron density of the 2DEG, and $m^*$ the electron effective mass.

Thus, from equations (13) and (14) the magnetoconductivities can be calculated, [Hidalgo 1995, 1998, 2007]. For testing and comparing our model we have used the experimental measurements by Dean et al., [2011]. In Figure 2 we show the results at high magnetic field (35 T) for the Hall magnetoconductivity, (a), and both magnetoresistivities, (b), as a function of the gate voltage, and for the series corresponding to the odd denominator $q$=3.

**Study of the experimental maxima of the SdH oscillations**

From the experimental SdH oscillations of any GFQHE observation, one can check that the maxima of every oscillation match with the values of the magnetic field given by the relation, [Hidalgo, 1995, 2013, 2014],

$$B_q \approx 3\frac{m^* E_F}{\hbar e} \qquad (16)$$

$E_F$ given by (15). Even more, each fraction $p$/3 corresponds to the plateau with the same fractional number.

**Summary and Conclusions**

We have presented a different approach to the GFQHE, being the change in the symmetry of the Hamiltonian of any electron of the 2DEG, when the magnetic field is increasing in the presence of the electrostatic potential due to the ionized impurities, the responsible of it. As the magnetic field increases the effect of that electrostatic potential evolves; changing in turn the spatial symmetry of the Hamiltonian: from continuous to discrete one. Therefore, in the case of the graphene the series of fractions obtained {1/3, 2/3, 1, 4/3, 5/3, 2, 7/3, 8/3, 3, 10/3, 11/3, 4, 15/3}. And this states are consequence of the appearance of a correlation among the cyclotron orbits with correlation index $\eta$=1, [Hidalgo, 2013].

Therefore the model justifies all the odd $q$=3 plateaux observed in the experiments on graphene. Besides, the present work shows that its physical origin is the same than this observed in QW semiconductors.

Thus, we hope that the shown approach could be a good starting point to understand the magnetotransport phenomena related to graphene.

**Acknowledgments:**

The author would like to thank R. Cangas for valuable discussions

**Table I: Fractional energy states for correlation length $\eta$=1.**

| $\dfrac{E}{\hbar\omega_0}$ | $\left(n-\dfrac{1}{q}\right)$ | | | | $\left(n+\dfrac{1}{q}\right)$ | | | | |
|---|---|---|---|---|---|---|---|---|---|
| | n=1 | n=2 | n=3 | n=4 | n=0 | n=1 | n=2 | n=3 | n=4 |
| q=3 | $\dfrac{2}{3}$ | $\dfrac{5}{3}$ | $\dfrac{8}{3}$ | $\dfrac{11}{3}$ | $\dfrac{1}{3}$ | $\dfrac{4}{3}$ | $\dfrac{7}{3}$ | $\dfrac{10}{3}$ | $\dfrac{13}{3}$ |

All the fractions shown in the table have already been observed in the experiments, [Bolotin et al., 2009], [Du et al., 2009], [Dean et al., 2011], [Gharari et al., 2011]. *n* corresponds to the Landau levels. The main odd denominator fraction usually observed in graphene is related to correlation length $\eta = 1$. (In Hidalgo [2013] can be found a detailed descripton of the model for the FQHE in QW semiconductors, where several series of the fractional plateaux are measured.)

**Figure legends:**

**Figure 1: Sketch with an artistic picture of the expected short-range order in a two-dimensional electron system due to an ionized impurities distribution underneath, and the arrangement of the corresponding electron cyclotron orbits**

Expected close packing arrangement of identical electron cyclotron orbits of a 2DEG in fractional quantum Hall conditions. In correspondence with Table I, the correlation detailed in the picture corresponds to a correlation length $\eta = 1$. (With black points are represented the impurities. $d_i$ is the mean distance among the ionized impurities.)

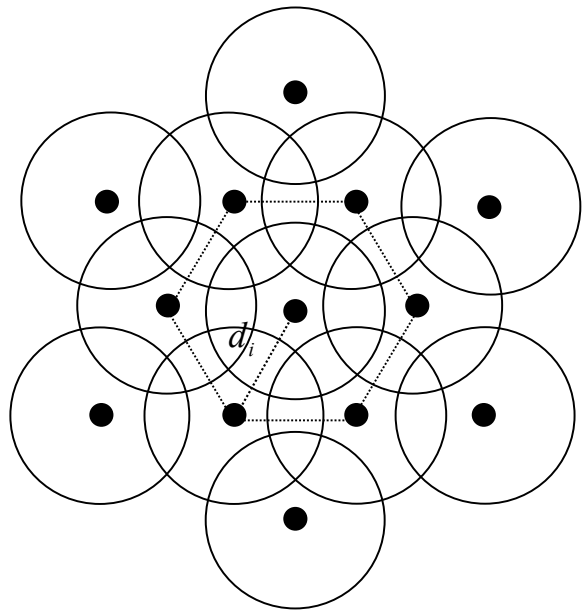

**Figure 2: Model's simulation of the fractional quantum Hall magnetoconductivity and both magnetoresistivities as a function of the gate voltage. It is shown the set of odd denominator states *q*=3, and the corresponding plateaux.**

We have simulated with our model the fractional quantum Hall effect in the conditions of the measurements by Dean et al. [2011]. The effective mass used in the simulation is 0.0125*m*, *T*=0.3 K, *g*=2, and $\Gamma$=1.8×10$^{-20}$ J. a) Hall magnetoconductivity where the plateaux have been highlighted. b) Both magnetoresistivities.

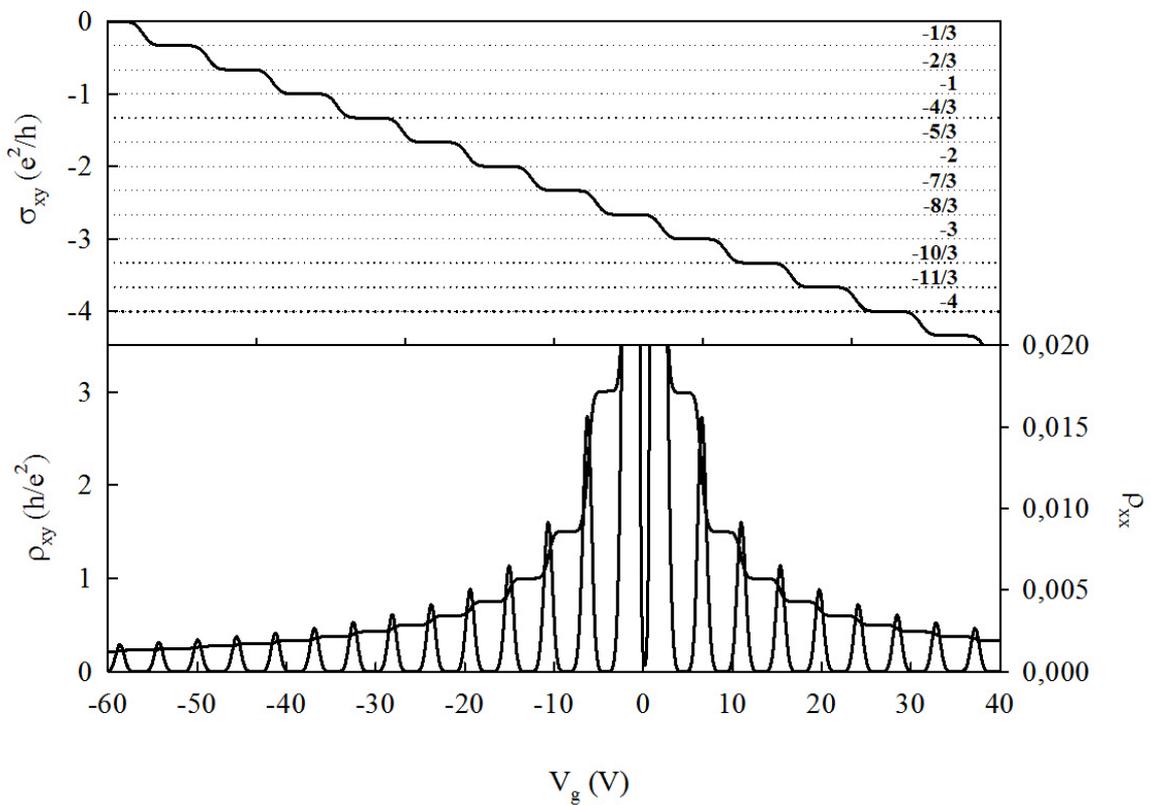